\documentclass{revtex4}   

\usepackage{amsmath}    
\usepackage{graphicx}   

\newcommand{\be}{\begin{equation}}
\newcommand{\ee}{\end{equation}}

\begin{document}

\title{Equation of state for hard sphere fluids
with and without Kac tails} 

\author{Emmanuel Trizac}
\affiliation{Universit\'e Paris-Sud, LPTMS, UMR 8626, Orsay Cedex, F-91405 and
CNRS, Orsay, F-91405}\author{Ignacio Pagonabarraga}
\affiliation{Departament de F\'\i sica Fonamental, Universitat de Barcelona,
Carrer Mart\`\i\ i Franqu\`es 1, 
E-08028 Barcelona}

\date{\today}

\begin{abstract}
In this note, we propose a simple derivation of the one dimensional hard rod
equation of state, with and without a Kac tail (appended long range and weak 
potential). 
The case of hard spheres in higher dimension is also
addressed and it is shown there that our arguments --which avoid any 
mathematical complication-- allow to recover the
virial form of the equation of state in a direct way.
\end{abstract}

\maketitle

\section{Introduction}


The equation of state is a central object in the statistical mechanics
description of matter. Bearing the signature of inter-particle
interactions and being easily measured experimentally, 
it establishes a connection between microscopic and macroscipic features
that long remained elusive.\cite{Brush} In particular,
many efforts in the 19th century aimed at explaining 
the deviations from the ideal gas law. Progresses in the 
field have been hindered by the positivist storm of criticisms directed 
against kinetic theory and the associated atomistic viewpoint.\cite{Brush}
Today, the machinery of statistical mechanics 
offers a consistent framework for those problems,\cite{Andrews,Chandler}
with the pedagogical drawback that 
very few interacting systems have an equation
of state that can be written in closed form.

The one dimensional hard-rod fluid (also called Tonks or Jepsen 
gas \cite{Tonks,Jepsen})
is one notable exception. 
For $N$ rods with a distribution 
of lengths $\{\ell_i\}_{1\leq i\leq N}$, enclosed in a line of total
length $L$, the pressure $P$ can simply be written \cite{Tonks}
\be
P \,=\, \frac{\rho k T }{1-\eta} ,
\label{eq:eos}
\ee
where $kT$ is the thermal energy, $\rho=N/L$ denotes the density,
and $\eta = \sum_i \ell_i/L$ is the line covering fraction
(so that $\eta/\rho$ is the mean rod length).
In addition, several equilibrium and non-equilibrium quantities
can be computed,\cite{Jepsen} and the model
provides a reference system for perturbative treatments, one of which 
being addressed below, with inclusion of a so-called Kac pair potential
of interaction between the particles.\cite{Kac1,Kac2}

In standard textbooks \cite{Andrews,Chandler} as well as in the original
papers,\cite{Tonks,Salz,Lebo,Reiss74} the derivation of Eq. (\ref{eq:eos}) 
is not straightforward, and it is 
our purpose here to propose an alternative concise argument relying on
simple physical considerations rather than partition functions and mathematical computations.
When transposed to systems of higher dimensions $d>1$, the argument
provides the exact equation of state of a hard sphere fluid, which
however cannot be written in closed form as an explicit function of 
density.\cite{Hansen}

\section{Equation of state of hard rod and hard sphere fluids}

We begin with the one dimensional case (hard rods). 
To achieve our goal of computing exactly the pressure, 
we consider several related ``Gedankenexperiments''.
We first make the remark that upon rescaling all lengths 
in the problem by a factor $1+\epsilon$
\be
\ell_i \to (1+\epsilon)\ell_i \hbox{ for all } 1\leq i\leq N, 
\quad L\to (1+\epsilon)L,
\label{eq:expansion}
\ee 
the reversible work $\delta W_{\text{total}}$
associated with this process reduces to its 
ideal gas contribution since excluded volume is irrelevant in this 
transformation (it is of course essential to rescale the ``box'' length $L$) ; 
only the ideal entropy of mixing is affected in this simple ``zoom''. 
We then have 
\be
\delta W_{\text{total}} = - \rho k T\, \delta L,
\label{eq:Wtotal}
\ee
where $\delta L = \epsilon L$ is the total length 
change. 
Equation (\ref{eq:Wtotal}) holds beyond hard body interactions
and applies whenever the problem at hand is governed by a unique
length scale.\cite{rque} It would in particular hold for Lennard-Jones
interactions.
For $\delta L>0$, the work $\delta W_{\text{total}}$
is negative.
If only the box length is expanded, the work 
$\delta W_{\text{total}}$ received by the gas is
obviously negative. The additional particle expansion considered here
amounts to a positive contribution to $\delta W_{\text{total}}$,
which turns out to be smaller in absolute value than the former
one. The resulting cost is negative, as the following discussion 
explicitly shows.

We now perform the expansion (\ref{eq:expansion}) in two steps and 
compute separately the corresponding reversible work required\\
\noindent a) First, rod sizes are slowly and sequentially rescaled 
(i.e. one at a time): 
$$\ell_1 \to (1+\epsilon)\ell_1, ~\ell_2 \to (1+\epsilon)\ell_2...$$ 
\noindent b) Second, the confining box size is expanded $L \to (1+\epsilon)L$.\\
\noindent It is essential to realize that in step a), any rod that is expanded
behaves as a confining wall (piston) which ``pushes'' the fluid so that
the work needed to rescale particle $i$ is 
$P \delta \ell_i = P \epsilon \ell_i$ (we comment further below this 
fundamental fact). 
Consequently, the work needed to
perform step a) reads 
\be
\delta W_a = \sum_{i=1}^N P \delta \ell_i = P \delta L_p,
\label{eq:Wa}
\ee
where $L_p$ is the total length occupied by the rods 
(here, $\delta L_p = \eta \delta L$). For step b), the work reduces to
\be
\delta W_b = -P \delta L.
\ee
Summing both contributions, we have 
$\delta W_{\text{total}}  =\delta W_a + \delta W_b$ and gathering results,
this implies 
\be
-\rho kT\, \delta L = P \,\eta \,\delta L - P \,\delta L \quad\Longrightarrow\quad
P \,=\, \frac{\rho k T }{1-\eta},
\label{eq:eos1d}
\ee
which is the correct result. We stress that irrespective of the sign of
the scaling parameter $\epsilon$, the moves considered in steps a) and b)
do not produce overlaps between particles nor between particles and confining
boundaries, in the same way as a moving boundary like a piston does not 
lead to overlaps. We also note that from the form of the equation of state
(\ref{eq:eos1d}), all virial coefficients are unity. This can be checked 
by explicit calculations.\cite{Bishop}

Following the same line of reasoning,
we  investigate the problem of
hard spheres in dimension $d>1$.
One still has
\be
\delta W_{\text{total}}  =\delta W_a + \delta W_b \quad \hbox{with} \quad
\delta W_{\text{total}} =-\rho kT \delta V, 
\label{eq:Wtot}
\ee
$V$ being now the volume
of the system, and $\delta W_b= -P \delta V$. However, $\delta W_a$
is not as straightforwardly related to the pressure as in (\ref{eq:Wa})
and to obtain its expression, we introduce the pair correlation function
$g(r)$.\cite{Chandler,Hansen} For the sake of simplicity, we restrict
to the monodisperse case where all particles have the same diameter $\sigma$.
The force per unit area felt by a given particle is 
of kinetic origin and reads $\rho kT g(\sigma)$~;
remembering that a particle is surrounded by an excluded volume sphere 
of diameter $2\sigma$ where no other particle's center of mass may be found,
we can write the work needed to expand $\sigma$ into 
$\sigma +  \delta \sigma$ as 
\be
\delta W_{\sigma \to \sigma +\delta \sigma}  = \rho kT \, g(\sigma) \, 
\delta V_{\text{sweep}},
\ee
where $\delta V_{\text{sweep}}$ is the volume change of the excluded volume
sphere. The latter quantity is related to the $d$-dimensional surface 
$S_d(2\sigma)$ of a sphere with radius $2\sigma$ through
\be
\delta V_{\text{sweep}} = S_d(2\sigma) \, \frac{\delta \sigma}{2}  = 
2^{d-1} S_d(\sigma) \,\frac{\delta \sigma}{2}.
\ee
Summing over all particles in the system, we get 
\be
\delta W_a = \sum_i \delta W_{\sigma \to \sigma +\delta \sigma} =
\rho kT g(\sigma) 2^{d-1} \, \delta V_p
\label{eq:200}
\ee
where, keeping previous notation, $\delta V_p$ is the change of the total volume
occupied by the spheres and therefore $\delta V_p/\delta V=\eta$,
the volume fraction.\cite{rque44}
Returning to Eq.(\ref{eq:Wtot}), we obtain
\be
-\rho kT = \rho k T \, 2^{d-1}\,\eta\, g(\sigma) -P,
\ee
and hence the equation of state
\be
\frac{P}{\rho kT} \,=\, 1 + 2^{d-1}\,\eta\, g(\sigma).
\label{eq:eoshs}
\ee
To our knowledge, the simplest derivation of this result involves
the virial theorem \cite {Hansen} 
and turns out to be more complicated and physically less
transparent.\cite{rque2}
As a by-product, comparing (\ref{eq:eoshs}) and (\ref{eq:eos1d})
in one dimension, we obtain that the pair correlation function at contact
for hard rods takes the value 
\be
g(\ell) \, = \, 1/(1-\eta),
\label{eq:g1d}
\ee
a standard result.\cite{Salz,Reiss74}
We also emphasize that it is straightforward to generalize 
(\ref{eq:200}) and (\ref{eq:eoshs}) to the polydisperse case with size distribution
$f(\sigma)$ (see e.g appendix B of Ref. \cite{Zhang}), with the result
\be
\frac{P}{\rho kT} -1=  \eta \int d\sigma d\sigma' f(\sigma) f(\sigma') 
g(\sigma/2+\sigma'/2) 
\frac{\sigma(\sigma+\sigma')^{d-1} }{\langle\sigma^d\rangle},
\ee
where $\langle\sigma^d\rangle = \int \sigma^d f(\sigma) d\sigma$.

\section{Inclusion of a Kac tail}

It proves instructive to consider also the case of particles interacting
with a so-called ``Kac tail'' since not only can the proper pressure
be recovered, but also some light be shed on the nature of the expansion
processes underlying our arguments. We assume that in addition to the
usual hard sphere term, particles interact 
with a very long range and weak pair potential so that the interaction potential
reads
\be
\phi(r) \,=\, \left\{ 
\begin{array}{lc}
\infty              & \hbox{for } r < \sigma \equiv \ell \\
\gamma \exp(-r/r^*) & \hbox{for } r > \sigma,
\end{array}
\right. 
\label{eq:Kacpot}
\ee
where the range $r^*$ is larger than any microscopic distance 
($\rho r^* \gg 1 $) and
the amplitude $\gamma$ is small. \cite{rque10} This quantity can be positive
(repulsive tail) or negative (attractive tail).
It has been shown 
--with all mathematical $i$-s dotted-- that the
corresponding equation of state is of the van der Waals form 
and reads \cite{Kac1}
\be
P \,=\, \frac{\rho k T }{1-\eta} + \alpha \, \rho^2,
\label{eq:Kac1}
\ee
where $\alpha = \gamma \,r^* $. This result has been generalized to
an arbitrary space dimension \cite{Kac2} in which case again, the
correction to the hard sphere pressure is simply $\alpha \rho^2$,
with now $\alpha = \int \phi({\bf r})\, d^d{\bf r}/2$, where the integral
runs outside the excluded volume sphere (i.e. $r>\sigma$). Actually, the
result holds irrespective of the precise form of the interaction 
potential,\cite{Andrews} as is also clear from the following argument.
We restrict to $d=1$, but space dimension is largely immaterial.
From $\delta W_{\text{total}}  =\delta W_a + \delta W_b$, we have
\be
(P-\rho kT) \delta L = \delta W_a,
\label{eq:aa}
\ee 
and our objective is now
to compute the latter term. In step a), it is understood that
the range of the potential $r^*$ is expanded, at fixed amplitude 
$\gamma$, to remain proportional to 
the rod size $\ell$: $\delta r^*/r^* = \delta \ell/\ell = \delta L/L$.
A key point in the analysis is that the tail $r>\ell$ of 
potential does not affect the relative positions of the particle from what they
would have as hard rods at the same density $\rho$.\cite{Kac2} This follows
from the fact that the amplitude $\gamma$ should be taken 
extremely small.\cite{rque10} The pair distribution function $g(r)$ 
is thus unaffected by the tail and we have in particular the contact 
value (\ref{eq:g1d}). 
When a given rod of size $\ell$ is expanded, 
the work necessary can be written as the sum of a kinetic contribution
(one has to ``push'' the particles that are in direct contact with the
particle of interest), and a (long range) contribution arising 
from the tail of the potential :
\be
\delta W_{\ell \to \ell +\delta \ell}  = \rho_{\text{contact}} kT \,\delta \ell \, 
+ \delta W_{\text{tail}} =  \, \frac{\rho kT}{1-\eta} \, \delta \ell\,
+ \delta W_{\text{tail}}.
\ee
Since any particle experiences an average potential energy $2 \gamma r^* \rho$,
as follows from integrating (\ref{eq:Kacpot}), we have 
$ \delta W_{\text{tail}} = \delta(2 \gamma r^* \rho) $ where the
variation is computed at fixed amplitude and density. Summing over all 
particles, we obtain
\be
\delta W_a \, = \frac{\rho kT}{1-\eta} \, \eta \, \delta L 
\, +\, \frac{1}{2} \, N \, 2 \gamma \rho \,\delta r^*,
\label{eq:bb}
\ee
where the factor $1/2$ corrects for double-counting. Remembering that
$\delta r^* / r^* = \delta L/L$, we arrive at
\be
\delta W_a \, = \frac{\rho kT}{1-\eta} \,  \eta \,\delta L 
\, +\,  \underbrace{\gamma  r^{\smash{*}}}_{\equiv \alpha} \,\rho^2 \, \delta L.
\ee
Inserting this result into (\ref{eq:aa}) yields Eq. (\ref{eq:Kac1}),
a result otherwise difficult to derive. \cite{Kac1,Kac2}
As is intuitively clear, Equation (\ref{eq:Kac1}) indicates that
repulsive interactions ($\gamma$ and $\alpha$ positive) 
lead to an enhanced pressure.

It seems appropriate at this point to emphasize that 
$\delta W_a \neq P \delta L_p$ where as above, $\delta L_p = \eta\delta L$
is the total length variation of the rods: expanding a given rod differs
from moving a piston in that the piston has to be held fixed, while
our move pertains to a given particle in the fluid, that is therefore free 
to move. If we would consider the particle expanded to be fixed at a given
location, and with a slowly increasing size ($\ell \to\ell + \delta \ell$),
the work required would reduce to $P \delta \ell$ since the particle 
would act as a piston. This work can again be expressed as the
sum of a kinetic pressure term {\em with the same value of the pair
distribution at contact as given by Eq. (\ref{eq:g1d})}, and
an energy potential difference:
\begin{subequations}
\begin{align}
P \delta \ell \,&= \, \rho_{\text{contact}} kT \,\delta \ell \, + \,
\delta (N \gamma \, r^* \, \rho) \\
&=\,\rho \, \frac{1}{1-\eta} kT \,\delta \ell \, + \,
\delta (N \gamma \, r^* \, \rho).
\end{align}
\end{subequations}
The difference with Eq. (\ref{eq:bb}) is that now the range $r^*$
is fixed but the density varies according to 
$\delta \rho/\rho = -\delta L/L = \delta \ell/L$. Hence, 
\be
P \delta \ell \,=\,\frac{\rho}{1-\eta} kT \,\delta \ell \,+\,
\gamma r^* \,\rho^2 \, \delta \ell,
\ee
and we recover Eq. (\ref{eq:Kac1}). In absence of the Kac tail
($\gamma=0$),
fixing or not the expanded particle is unimportant and the fact that we
there have $\delta W_a = P \delta L_p$ can be viewed as a consequence
of the hard potential used (with interactions at contact only
and a pressure that only depends --apart from $\rho$ and $T$--
on the pair distribution function 
at contact)

\section{Concluding remarks}

We have proposed an exact derivation of the equation of state 
of certain simple liquids: hard rod systems with and without a long range
pair potential of interaction (Kac tail). The argument --based
on scaling considerations-- can 
be easily extended to hard core particles in higher dimensions
(hard discs, hard spheres\ldots). The interest of the approach lies
in its simplicity and it is our hope that the method put forward here
is instructive and may be useful to illustrate an advanced
undergraduate statistical mechanics course. 

To our knowledge, pegagogical accounts avoiding 
mathematical complication on the present topics are scarce
in the literature. We note however that a relatively simple
approach generalizing the original Bernouilli derivation 
of the ideal gas equation of state has been proposed.\cite{DelRio} 
In addition to being heuristic, this method could not
be transposed to the case of Kac tails.

Finally, we note that our arguments bear some similarities with 
the scaled particle theory developed in the 1960s,
\cite{Reiss,Lebo,Reiss74,Hansen} but are nevertheless
distinct. Scaled particle theory is based on the computation of the
reversible work needed to create a spherical cavity from which the
centers of other spheres are excluded. This expression is then related
to the density of particles at contact with the cavity boundary.
This can be achieved exactly in one dimension, since there are then
no curvature effects. The latter remark also explains why we have
been able to derive the pressure in closed form for $d=1$.

{\em Acknowledgements} We would like to thank Andres Santos for 
useful comments.



\begin{thebibliography}{10}
\bibitem{rque}
To be more specific, 
this means that introducing reduced coordinates 
$\widetilde x_i = x_i/L$ and reduced lengths $\widetilde \ell_i=\ell_i/L$, 
the Hamiltonian expressed 
in terms of $\{ \widetilde x_i, \widetilde \ell_i\}$ should be 
$L$-independent so that the canonical partition function reads
\be
Z(T,L,N,\{\ell_i\}_{1\leq i\leq N}) = L^N \widetilde 
Z(T,\{\widetilde\ell_i\}_{1\leq i\leq N}).
\ee
For other 
interactions than hard core, the above scaling relation may still hold,
which immediately leads to (\ref{eq:Wtotal}).
\bibitem{rque44}
The fundamental difference between $d=1$
and higher space dimensions is that one has $P = \rho kT g(\sigma)$
in one dimension, since the particle-particle pair distribution 
function at contact then coincides with $g_{pw}(\sigma)$, the particle-wall distribution 
function at contact. For $d>1$, one has $P = \rho kT g_{pw}(\sigma)$, which
differs from $\rho kT g(\sigma)$.
\bibitem{rque10}
Formally, the limit $r^* \to \infty$, $\gamma \to 0$
with $\gamma \,r^*$ fixed is required, and should be taken after the
thermodynamic limit. \cite{Kac1,Kac2}
\end{thebibliography}
\end{document}